\documentclass[prd,floats,preprint,superscriptaddress,eqsecnum,floatfix,nofootinbib,12pt]{revtex4}

\usepackage{amsmath}
\usepackage{graphics, graphicx}
\usepackage{epsfig}
\usepackage{ulem}
\usepackage{color}

\renewcommand{\Im}{{\rm Im}}
\renewcommand{\Re}{{\rm Re}}

\def\phih{\hat\phi}

\def\cP{{\cal P}}

\def\1p{{(1p)}}
\def\be{\begin{equation}}
\def\ee{\end{equation}}
\def\beq{\begin{eqnarray}}
\def\eeq{\end{eqnarray}}
\def\cf{}

\def\p0{\phi_0}
\def\z0{\zeta_0}

\def\uf{}
\def\vf{}

\def\wf{}

\def\zf{}
\def\af{}
\def\vf{}
\def\uf{}
\def\qf{}
\def\clr{}

\def\vx{{\vec x}}
\def\lm{\lambda_{-}}
\def\lp{\lambda_{+}}
\def\th{{\tilde h}}
\def\cP{{\cal P}}
\def\cK{{\cal K}}
\def\lt{{left }}
\def\rt{right }

\newcommand{\ttle}[1]{{\it #1}}

\begin{document}

\vspace{1cm}

\title{Holographic No-Boundary Measure}

\author{Thomas Hertog}
\affiliation{APC, UMR 7164, 10 rue A.Domon et L.Duquet, 75205 Paris, France {\it and}\\
Institute for Theoretical Physics, University of Leuven, 3001 Leuven, Belgium}
\author{James  Hartle}
\affiliation{Department of Physics, University of California, Santa Barbara,  93106, USA}

\bibliographystyle{unsrt}

\vspace{1cm}

\begin{abstract}

We show that the complex saddle points of the no-boundary wave function with a positive cosmological constant and a positive scalar potential have a representation in which the geometry consists of a regular Euclidean AdS domain wall that makes a smooth transition to a Lorentzian, inflationary universe that is asymptotically deSitter. The transition region between AdS and dS regulates the volume divergences of the AdS action and accounts for the phases that explain the classical behavior of the final configuration. This leads to a dual formulation in which the semiclassical no-boundary measure is given in terms of the partition function of field theories on the final boundary that are certain relevant deformations of the CFTs that occur in AdS/CFT. We conjecture that the resulting dS/CFT duality holds also beyond the leading order approximation.

\end{abstract}

\vskip.8in
\vspace{1cm}


\maketitle

\section{Introduction}
\label{intro}

In cosmology one is interested in computing the probability measure for different classical configurations {\af of geometry and fields} on a spacelike surface $\Sigma$. This measure is given by the universe's quantum state. In a series of papers  \cite{HHH08,HHH10a} we have calculated the tree level measure predicted by the no-boundary wave function {\wf (NBWF)} \cite{HH83} for gravity coupled to a {\af positive cosmological constant and}  a scalar field with a positive potential. {\af Predictions for our observations are obtained by further conditioning on our observational situation and its possible location in each history, and then summing {\clr over} what is unobserved \cite{HHH10b}.}

The {\af formal} sum over geometries {\af usually used to specify} the NBWF \cite{HH83} {\af is} difficult to define precisely. Further, the indefiniteness of the Euclidean gravitational action requires a conformal factor rotation \cite{Gibbons78} of the sum whose exact nature has not been specified beyond the semiclassical level \cite{HH90}. It is therefore of interest to find a mathematically more precise formulation of the NBWF that allows one to reliably calculate the probability measure beyond the saddle point approximation. 

In this paper we initiate a novel approach to this problem that aims to formulate the NBWF in string theory. In particular we show that one can use the Euclidean AdS/CFT correspondence\footnote{See \cite{Maldacena03,Horowitz04,Maldacena11} for discussions of AdS/CFT in the context of the wave function of the universe for $\Lambda<0$.} \cite{Maldacena98,Witten98,Gubser98} to derive a dual formulation of the semiclassical\footnote{\af By `semiclassical' we mean leading order in $\hbar$ {\uf and $\alpha'$}.} NBWF in terms of the partition function of a Euclidean field theory defined on the future boundary conformal to $\Sigma$. In the spirit of AdS/CFT we conjecture that the duality extends beyond the leading order approximation. The dual field theory description would then give a precise meaning (in a certain limit) to the notion of a wave function of the universe in the context of string theory and provide a new method to compute the string and quantum corrections to the tree level no-boundary measure\footnote{This can also be viewed as a novel application of AdS/CFT to cosmology, or as a realization of dS/CFT.}.

In its current form the NBWF is defined by a sum over regular complex four-geometries {\af and four-dimensional matter field configurations on a four-disk $M$ with boundary $\Sigma$. The configurations are } weighted by $\exp(-2I/\hbar)$ {\wf where $I$ is 
the Euclidean action of geometry and field. {\af In this paper we take this to be the action of }Einstein gravity coupled to a positive cosmological constant and scalar matter fields with a positive potential\footnote{The same {\af results} can be established starting from a negative cosmological constant and a negative scalar potential \cite{HH11}.}. In the semiclassical approximation {\wf the NBWF} predicts an {\it ensemble} of classical, Lorentzian universes. Each member in this ensemble is associated with a complex saddle point geometry, which is an extremum of $I$ that is regular {\af on $M$} and matches onto the classical {\af configuration} on $\Sigma$.

To leading order in $\hbar$ the probabilities of different classical configurations are proportional to $\exp(-2I_R/\hbar)$, where $I_R$ is the real part of the action of the corresponding saddle point. These probabilities are conserved along a classical trajectory as a consequence of the Wheeler-DeWitt equation \cite{HHH08} and therefore constitute the tree-level no-boundary probability measure on an ensemble of coarse-grained\footnote{{\uf In the expanding branch of the NBWF one finds that the ensemble of classical histories predicted by the NBWF on a surface $\Sigma$ is a coarse-graining of the ensemble predicted on a larger surface $\Sigma'$. This is expected on general grounds, because the wave function at $\Sigma$ automatically coarse-grains over all bifurcations {\clr of histories} to the future of $\Sigma$.}} classical {\it histories}. 

The action of a saddle point is an integral of its complex geometry and fields that includes an integral over  time. Different complex contours for this {\clr time}  integral give different representations of the saddle point, {\af each giving the same probability for the classical configuration the saddle point} corresponds to. This freedom in the choice of contour gives physical meaning to a process of analytic continuation --- not of the Lorentzian classical histories themselves --- but of the saddle points that define their probabilities.   

Using this freedom of choice of contour, we {\af identify two different useful representations of a saddle point corresponding to an asymptotically deSitter, classical, Lorentzian history.  In one representation (dS) the interior geometry behaves as though the cosmological constant and the scalar potential were positive. In the other (AdS) the Euclidean part of the interior geometry behaves as though these quantities were negative, defining a regular, asympotically AdS domain wall. Asymptotically Lorentzian deSitter (dS) universes and Euclidean anti-de Sitter (AdS) spaces are thereby connected by the NBWF. } 

{\af We find that the action $I$ of the saddle points can be expressed schematically as 
\begin{equation}
I = I_{\rm dS} = I^{\rm  reg}_{\rm DW} + iS_{ct} .
\label{scnbwf-schematic}
\end{equation}
Here $I_{\rm dS}$ is the action in the deSitter representation and $I^{\rm  reg}_{\rm DW}$ is the regularized action of the Euclidean domain wall. $S_{ct}$ is a real surface term. 
Eq \eqref{scnbwf-schematic} (expressed precisely in \eqref{scnbwf}) is our core result. 
It implies that the requirement that a configuration on $\Sigma$ behaves classically, with constant $I_R$, automatically regulates the volume divergences associated with the action of the Euclidean AdS regime of the saddle point. Furthermore, it implies that the leading order in $\hbar$ probabilities of classical, Lorentzian, asymptotically de Sitter histories can be calculated either from the dS representation {\qf of the saddle points or from their representation as Euclidean, asymptotically AdS, domain walls.} 

In a rather large class of models one can use the Euclidean AdS/CFT duality to replace the AdS domain wall factor $I^{\rm  reg}_{\rm DW}$ by {\qf minus the logarithm} of the partition function of a dual field theory. This leads to a dual formulation of the {\af semiclassical} NBWF -- and hence a concrete realization of a dS/CFT duality -- in terms of one of the known, unitary dual field theories familiar from AdS/CFT defined on the boundary conformal to $\Sigma$. In this dual description, the argument of the wave function (modulo the scale factor) enters as an external source in the dual partition function. The dependence of the {\clr partition function} on the values of these gives a dual no-boundary probability measure on the space of classical, inflationary histories.}

The resulting dS/CFT duality has several properties that have been conjectured or studied elsewhere (see e.g. \cite{Strominger01b,Strominger01,Maldacena03,Garriga09,Fadden10,Bousso10,Harlow11,Maldacena11,Anninos11}). 
In particular evolution in time of the universe, which corresponds to radial evolution in the saddle points, emerges in our framework as inverse RG flow in the dual as originally conjectured in \cite{Strominger01b}. On the other hand, an important distinction between our approach and most of the previous discussions of dS/CFT is that the gauge/gravity duality established here involves the quantum state of the universe. {\af In particular,} our setup relies neither on the analytic continuation of Lorentzian solutions in the complex plane, nor on a correspondence between (Lorentzian) dS and (Euclidean) AdS theories that involves the continuation of the dS radius and the time coordinate (or an analogous continuation in the dual parameters). Instead the connection between AdS and dS emerges at the level of the quantum state. This also means that even though the dual field theory lives on the future boundary of deSitter, it encodes the `initial' state of the universe and is thus closely connected to the resolution of the singularity problem in cosmology\footnote{\vf See the discussion of singularity resolution in the NBWF {\clr context}  in \cite{HHH08}.}.

The paper is organized as follows:  Section \ref{noboundary} reviews very briefly the NBWF and how it predicts Lorentzian histories. The details can be found in the series of papers cited above, especially \cite{HHH08}.  Section \ref{rep} discusses  the different possible representations of the complex saddle points of the NBWF, in particular the representation in which the Euclidean region exhibits an AdS geometry. Section \ref{nomatter} illustrates this explicitly in the simplest possible case --- empty deSitter space.  Section \ref{nonlinear} considers the ensemble of homogeneous  saddle points predicted by the NBWF for gravity coupled to a scalar field with a positive potential. These correspond to inflating asymptotically dS universes. Section \ref{more_general} discusses general, inhomogeneous saddle point configurations and their AdS representation. Finally in Section \ref{dualities} we derive a dual formulation of the NBWF by applying AdS/CFT to the general bulk geometries.

\section{Classical Predictions of the No-Boundary Quantum State}
\label{noboundary}
\subsection{The No-Boundary Wave Function of the Universe}
\label{nbwf}

A quantum state of the universe is specified by a wave function $\Psi$ on the superspace of  3-geometries and matter field configurations on a closed spacelike surface $\Sigma$.  Representing 3-geometries by metrics $h_{ij}(\vec x)$, and taking a single scalar field $\chi(\vec x)$ for the matter, we write (schematically) $\Psi=\Psi[h,\chi]$. 

We assume the no-boundary wave function (NBWF) as a model of the state \cite{HH83}. The NBWF is given by a sum over geometries $g$ and fields $\phi$ on a four-manifold $M$ with one boundary $\Sigma$. The contributing histories match the values $(h,\chi)$ on $\Sigma$ and are otherwise regular. They are weighted by $\exp(-I/\hbar)$ where $I[g,\phi]$ is the Euclidean action.  Schematically, 
\begin{equation}  
\Psi[h(\vec x),\chi (\vec x)] \equiv  \int_{\cal C} \delta g \delta \phi \exp(-I[g(x),\phi(x)]/\hbar) .
\label{nbwf}
\end{equation}
Here, $g(x)$ (short for $g_{\alpha\beta}(x^\gamma)$) and $\phi(x)$ are the histories of the 4-geometry and matter field. We take the Euclidean action $I[g(x),\phi(x)]$ to be a sum of the Einstein-Hilbert action (in Planck units where $\hbar=c=G=1$)
\begin{equation}
I_C[g] = -\frac{1}{16\pi}\int_M d^4 x (g)^{1/2}(R-2\Lambda) -\frac{1}{8\pi}\int_M d^4 x (h)^{1/2}K
\label{curvact}
\end{equation}
and the matter action 
\begin{equation}
I_{\phi}[g,\Phi]=\frac{3}{4\pi} \int_M d^4x (g)^{1/2}[(\nabla\phi)^2 +V(\phi)] 
\label{mattact}
\end{equation}
(where the normalization of the scalar field $\phi$ has been chosen to simplify subsequent equations and maintain consistency with our earlier papers.)
The integration in \eqref{nbwf} is carried out along a suitable complex contour which ensures the convergence of \eqref{nbwf} and the reality of the result  \cite{HH90}. 

In this paper we concentrate on models in which the cosmological constant $\Lambda$ and the potential $V$ in the action \eqref{curvact}-\eqref{mattact} are positive.
This means that with a positive signature convention {\clr for the metric}  the Euclidean action $I = I_C+I_{\phi}$ in \eqref{nbwf} is that of Einstein gravity coupled to a positive cosmological constant and a positive potential. With a negative signature convention this is minus the action\footnote{The simple relation under a change of signature between the action of dS gravity and AdS gravity coupled to scalar matter extends to four-dimensional supergravity, where the dS theory inherits a pseudo-supersymmetry from the supersymmetry of the AdS theory \cite{Cvetic95,Skenderis07}. It is sometimes referred to as a domain wall/cosmology correspondence. We note, however, that in our framework this does not arise as a correspondence between (solutions of) different theories, but rather as a property of the quantum state of a given theory.} of Einstein gravity coupled to a negative cosmological constant $-\Lambda$ and a negative potential $-V$.
Any notion of signature is of course meaningless for the complex metrics contributing to  \eqref{nbwf}. However the signature of the real boundary metrics $h$ is well-defined and  a convention for it} should be specified.  We take this to be positive in this paper. On the boundary $\Sigma$, $g(x)$ must induce $h(x)$. Hence the signature adopted for the boundary metrics determines which metrics $g(x)$ contribute to \eqref{nbwf}, hereby completing the definition of the NBWF. 

\subsection{Prediction of an Ensemble of Classical Histories}
\label{classens}

In some regions of superspace the path integral \eqref{nbwf} defining the NBWF can be approximated by the method of steepest descents. Then the NBWF will be approximately given by a sum of terms of the form 
\begin{equation}
\Psi[h,\chi] \approx  \exp\{(-I_R[h,\chi] +i S[h,\chi])/\hbar\} ,
\label{semiclass}
\end{equation}
one term for each complex saddle point (extremum).  Here $I_R[h,\chi]$ and $-S[h,\chi]$ are the real and imaginary parts of the Euclidean action, evaluated at the saddle point. 

In regions of superspace where, with an appropriate coarse-graining, $S$ varies rapidly compared to $I_R$ (as measured by quantitative classicality conditions \cite{HHH08}) the NBWF predicts that coarse-grained histories of geometry and fields behave classically. That is,  with high probability they exhibit  patterns of correlations in time summarized by classical deterministic laws. This is analogous to the prediction of the classical behavior of a particle in a WKB state in non-relativistic quantum mechanics. {\wf We therefore call points in superspace where the classicality conditions are satisfied `classical configurations'.} More specifically the NBWF predicts an ensemble of spatially closed, classical Lorentzian cosmological histories that are the integral curves of $S$ in superspace.  Integral curves are defined by integrating the classical relations  relating momenta $\pi_{ij}(\vec x)$ and $\pi_\chi({\vec x})$ to derivatives of the action 
\begin{equation}
\label{momenta}
\pi_{ij}(\vec x) =\delta S /\delta h_{ij}(\vec x), \quad \pi_\chi(\vec x)= \delta S /\delta \chi(\vec x) .
\end{equation}
The momenta are proportional to the time derivatives of $h_{ij}$ and $\chi$ {\wf so that equations \eqref{momenta} become differential equations for classical trajectories}. The solutions $h_{ij}({\vec x},t)$ and $\chi({\vec x},t)$ define field histories  ${\hat\phi}(x,t)\equiv\chi(x,t)$ and Lorentzian four-geometries $\hat g_{\alpha\beta}(x,t)$  by
\begin{equation}
\label{lorhist}
ds^2 = -dt^2 + h_{ij}(x,t)dx^i dx^j \equiv {\hat g}_{\alpha\beta}(x,t) dx^\alpha dx^\beta
\end{equation} 
in a simple choice of gauge\footnote{We follow the notation introduced in \cite{HHH08} that the complex extrema are denoted by $(g_{\alpha\beta}(x,t),\phi(x,t))$  and the real four-dimensional Lorentzian histories  by $({\hat g}_{\alpha\beta}(x,t),\phih(x,t))$. Occasionally, as above, when we want to emphasize that the Lorentzian histories are integral curves in superspace we will use the notation $(h_{ij}(x,t),\chi(x,t))$ for them.}. The real Lorentzian histories are therefore not the same as the complex saddle points that determine their probabilities. Further, the relations between superspace coordinates and momenta \eqref{momenta} mean that to leading order in $\hbar$,  and at any one time, the predicted classical histories do not fill classical phase space. Rather, they lie on a surface within classical phase space of half its dimension. 

Classicality in general requires appropriate coarse graining, that is, summing amplitudes over a bundle of nearby histories \cite{GH07,Har09,Har95c}. That is necessary {\wf both for decoherence} and to enable the destructive interference that suppresses amplitudes for non-classical coarse-grained histories. The validity of the WKB prescription given here depends on that coarse graining. The functions $h_{ij}(t,\vx)$ and $\chi(t,\vx)$ should be understood as labels for coarse grained histories. Only their structure on scales larger than the coarse graining {\uf scale ${\zeta}$} is {\uf classically predicted and} relevant for distinguishing one coarse-grained history from another. 

{\uf It turns out that in the inflationary universes predicted by the NBWF, the coarse-graining scale required for classicality on a surface $\Sigma$ is essentially its horizon size. Hence the coarse-graining scale on $\Sigma$ is larger than the coarse-graining scale on a surface $\Sigma'$ at a larger scale factor. Thus the ensemble of classical histories obtained from the set of classical configurations on $\Sigma$ by integrating \eqref{momenta} will be a coarse graining of the classical ensemble predicted by the NBWF evaluated on $\Sigma'$.}

{\uf Each individual coarse-grained classical history of a classical ensemble} has a probability\footnote{In the terminology used in our other papers these were called bottom-up probabilities to distinguish them from top-down probabilities that are conditioned on (part of) our data. Top-down probabilities are relevant for predicting our observations. Bottom-up probabilities are relevant for discussing the probabilities of features of  the universe whether or not there are any observers. All probabilities here are bottom-up.} proportional to $\exp[-2 I_R(h,\chi)/\hbar]$ to leading order in $\hbar$. The probabilities $\exp[-2 I_R(h,\chi)/\hbar]$ are constant along the integral curves given by \eqref{momenta} as a consequence of the Wheeler-DeWitt equation (cf \cite{HHH08}). Hence these give the tree level no-boundary measure of different possible universes in {\uf a classical ensemble} predicted by the NBWF.

\section{Representations of Complex Saddle Points}
\label{rep}

\subsection{Homogeneous and Isotropic Saddle Points}
\label{homoisobkgnds}
We begin by considering ${\cal O}(4)$ invariant saddle points of the NBWF for Einstein gravity coupled to a single scalar field $\phi$ moving in a positive potential $V(\phi)$. We assume that the cosmological constant term $\Lambda \equiv 3H^2$ is positive. In Section \ref{more_general} we will generalize our analysis to general inhomogeneous configurations on $\Sigma$.

The line element of a homogeneous and isotropic closed three-geomety is  
\begin{equation}
d\Sigma^2 = h_{ij} dx^i dx^j = b^2 \gamma_{ij}(x^k) dx^i dx^j
\label{homoiso3metric}
\end{equation} 
where $b$ is a constant {\cf (positive)} scale factor and $\gamma_{ij}$ is the metric on a unit round three sphere. Homogeneous and isotropic minisuperspace is therefore spanned by $b$ and the homogeneous value of the scalar field $\chi$. Thus $\Psi=\Psi(b,\chi)$. 

The line element of homogeneous and isotropic four-geometries on the manifold $M$ with one boundary  that are summed over in \eqref{nbwf} can be written
\begin{equation}
ds^2=N^2(\lambda) d\lambda^2 +a^2(\lambda) d\Omega^2_3 
\label{eucmetric}
\end{equation} 
where $(\lambda, x^i)$ are four real coordinates on the real manifold $M$.  Conventionally we take $\lambda=0$ to locate the center of ${\cal O}(4)$ symmetry (the South Pole (SP))  and $\lambda=1$ to locate the boundary of $M$ where histories match $(b,\chi)$.  Saddle points may be represented by complex metrics --- complex $N$ and $a$ --- but the coordinates  $(\lambda, x^i)$ are always real.  The Euclidean action is then\footnote{The scalar field has been rescaled from a conventional value by a factor to simplify the expression for the action as in \cite{HHH08}. However we have not rescaled the metic by a factor of $H^2$ as in that paper so that the cosmological constant is represented explicitly.} 
\begin{equation} 
I[a(\lambda),\phi(\lambda)] = \frac{3\pi}{4}\int^1_0 d\lambda N\Bigg\{ -a \left(\frac{a'}{N}\right)^2 -a +H^2a^3
+a^3\left[\left(\frac{\phi'}{N}\right)^2 + 2V(\phi)\right]\Bigg\}. 
\label{eucact}
\end{equation}
Equations sufficient for calculating the saddle points of the action $I$ are
\begin{subequations}
\label{euceqns_N}
\begin{equation}
\left(\frac{a'}{N}\right)^2 -1 +H^2a^2 +a^2\left[-\left(\frac{\phi'}{N}\right)^2 + 2 V(\phi)\right]=0,
\label{eucconstraint_N}
\end{equation}
\begin{equation}
\frac{1}{a^3N}\left(a^3\frac{\phi'}{N}\right)' - \frac{d V}{d\phi}= 0 , 
\label{eucphieqn_N}
\end{equation}
\end{subequations}
where a $'$ denotes a derivative with respect to $\lambda$.  The pair of functions $(a(\lambda), \phi(\lambda))$ defining saddle points contributing to the semiclassical wave function \eqref{semiclass} must be regular at $\lambda=0$ and match {\wf the real values} $(b,\chi)$ at $\lambda=1$. They will generally be complex --- fuzzy instantons.

\subsection{Different Representations of Complex Saddle Points}
The equations \eqref{euceqns_N} can be solved for $a(\lambda),\phi(\lambda)$ for any complex $N(\lambda)$ that is specified. Different choices of $N(\lambda)$ therefore give different representations of the same saddle point. (For real metrics of the form \eqref{eucmetric} different choices of $N(\lambda)$ are connected by coordinate transformations.)  A convenient way to exhibit these different representations is to introduce the function $\tau(\lambda)$ defined by
\begin{equation}
\tau(\lambda) \equiv \int_0^\lambda d\lambda' N(\lambda').
\label{deftau}
\end{equation} 
Different choices of $N(\lambda)$ correspond to different contours in the complex $\tau$-plane. Contours start from the SP at $\lambda=\tau=0$ and end at the boundary $\lambda=1$ with $\tau(1)\equiv \upsilon$. Conversely, for any contour $\tau(\lambda)$ there is an $N(\lambda)\equiv d\tau(\lambda)/d\lambda$. Each contour connecting $\tau=0$ to $\tau=\upsilon$ is therefore a different representation of the same complex saddle point. 

By using $d\tau=N(\lambda)d\lambda$ the saddle point equations \eqref{euceqns_N} can be written in the more compact form 
\begin{subequations}
\label{euceqns}
\begin{equation}
{\dot a}^2 -1 +H^2a^2 +a^2\left(-{\dot \phi}^2 + 2 V(\phi)\right)=0,
\label{eucconstraint}
\end{equation}
\begin{equation}
\ddot\phi + 3({\dot a}/a)\dot\phi  - \frac{d V}{d\phi}= 0 , 
\label{eucphieqn}
\end{equation}
\end{subequations}
where a dot denotes a derivative with respect to $\tau$. Solutions define functions $a(\tau)$ and $\phi(\tau)$ in the complex $\tau$-plane. A contour $C(0,\upsilon)$  representing a saddle point  connects the SP at 
$\tau=0$ to a point $\upsilon$  where $a(\upsilon)$ and $\phi(\upsilon)$ take the real values $b$ and $\chi$ respectively. For any such contour the action is given by  
\begin{equation} 
I(b,\chi) = \frac{3\pi}{4}\int_{{C}(0,\upsilon)}  d\tau \Big[ -a {\dot a}^2 -a +H^2a^3 +a^3\left({\dot\phi}^2 + 2V(\phi)\right)\Big] .
\label{eucact_tau}
\end{equation}
Assuming analyticity the result will be the same for any contour connecting the two points. Substituting \eqref{eucconstraint} in this expression we find a useful alternative expression of the action.
\be
I(b,\chi) = \frac{3\pi}{2}\int_{{C}(0,\upsilon)} d\tau a \left[ a^2\left(H^2+ 2V(\phi)\right)-1\right].
\label{eucact_alt}
\ee

We will refer to any solution $(a(\tau),\phi(\tau))$ of the equations \eqref{euceqns} satisfying the NBWF conditions of regularity at the SP as a `saddle point'.  Particular saddle points have particular uses. The saddle points contributing to the semiclassical approximation  of the NBWF evaluated at $(b,\chi)$  have points $\tau=\upsilon$ where $(a(\upsilon),\phi(\upsilon))$ take the real values  $(b,\chi)$. A saddle point that has a curve in the complex $\tau$-plane along which $(a(\upsilon),\phi(\upsilon))$ have real values is associated with a real history. A saddle point where the classicality conditions are satisfied in addition corresponds to a real Lorentzian history. Thus when we refer to a deSitter saddle point we mean one where there is such a curve with the geometry of deSitter space, etc.

Distorting one contour representing a saddle point into another representing the same saddle point provides a natural and general notion of analytic continuation. We emphasize this is not a continuation of  the Lorentzian histories into the complex plane. Neither is it a continuation between solutions of one theory to solutions of a different theory\footnote{In particular this does not involve a change in the contour defining the path integral \eqref{nbwf}.}. Rather it is a continuation of the complex saddle points that represent the Lorentzian histories in the NBWF {\clr and supply their probabilities} for {\clr one} given theory.   It is thus not a further assumption but a connection that is automatically available in this framework. We now illustrate this in a simple model.

\section{A Simple Model}
\label{nomatter}
\subsection{The No-Matter Model}
The case of no-matter field provides an example that is  oversimplified but nevertheless instructive because it is explicitly soluble. When $\phi=0$, or when the matter fields are absent from the Lagrangian altogether, the unique solution of \eqref{eucconstraint}  that is regular at $\tau=0$ is 
\be
a(\tau)=\frac{1}{H} \sin (H\tau) \ .
\label{defa}
\ee

Since $a(\tau)$ is an entire function, the action at an endpoint $\upsilon$ can be evaluated by doing the integral \eqref{eucact_tau}  along any contour $C(0,\upsilon)$ connecting $\tau=0$ to $\tau=\upsilon$. The result is 
\be
I(\upsilon) = -\frac{\pi}{2 H^2}[1-\cos^3(H\upsilon)] \  .
\label{Iups}
\ee

\subsection{ The deSitter Saddle Point}

Saddle points contributing to the semiclassical approximation of the NBWF have an endpoint $\upsilon$ where $a(\upsilon)$ takes a real, {\cf positive} value $b(\upsilon)$ [cf. \eqref{homoiso3metric}]. The most relevant case\footnote{The solution to the saddle point equations $a(\tau)=H^{-1}\sin(H\tau)=H^{-1}\sin[H(x+iy)]$ is real analytic, symmetric under $\tau \to -\tau$, and periodic in $x$ with period $2\pi/H$. There are therefore other values of $\upsilon$ where $a(\tau)$ is real besides those on the curve $x=\pi/2H$ analyzed above. These were discussed in Appendix A.1 of \cite{HHH08}. They are either solutions that are not regular on $M$, endpoints for which the classicality conditions are not satisfied, or ones where the classical history is the time reversed of the one discussed here.} is when the endpoint is located along the line $x=\pi/2H$. Along this line $a(y)$ takes the real values $(1/H)\cosh(Hy)$.  The action \eqref{Iups} at an endpoint $\upsilon=\pi/2H+iy_{\upsilon}$ has real and imaginary parts
\begin{subequations}
\label{actRandI}
\be
I(\upsilon)=I_R(\upsilon)-iS(\upsilon) \ , 
\label{Irealim}
\ee
with
\begin{equation}
I_R(\upsilon)= -\frac{\pi}{2H^2},  \quad 
S(\upsilon)=  -\frac{\pi}{2H^2} \sinh^3(Hy_{\upsilon}). 
\label{dS}
\end{equation}
\end{subequations}

For large $y_{\upsilon}$ the classicality condition is satisfied for this saddle point. The action $S$ varies rapidly when compared with $I_R$ because $I_R$ does not vary at all. A classical ensemble is predicted with one  classical history which is Lorentzian deSitter space. A discussion of probability is trivial when there is only one history in the sample space. But the weight  $\exp(-2I_R/\hbar)$ that will play this role in the more general context discussed below is the usual one for deSitter space. We call this the deSitter (dS) saddle point. 
\begin{figure*}[t]
\includegraphics[width=3.0in]{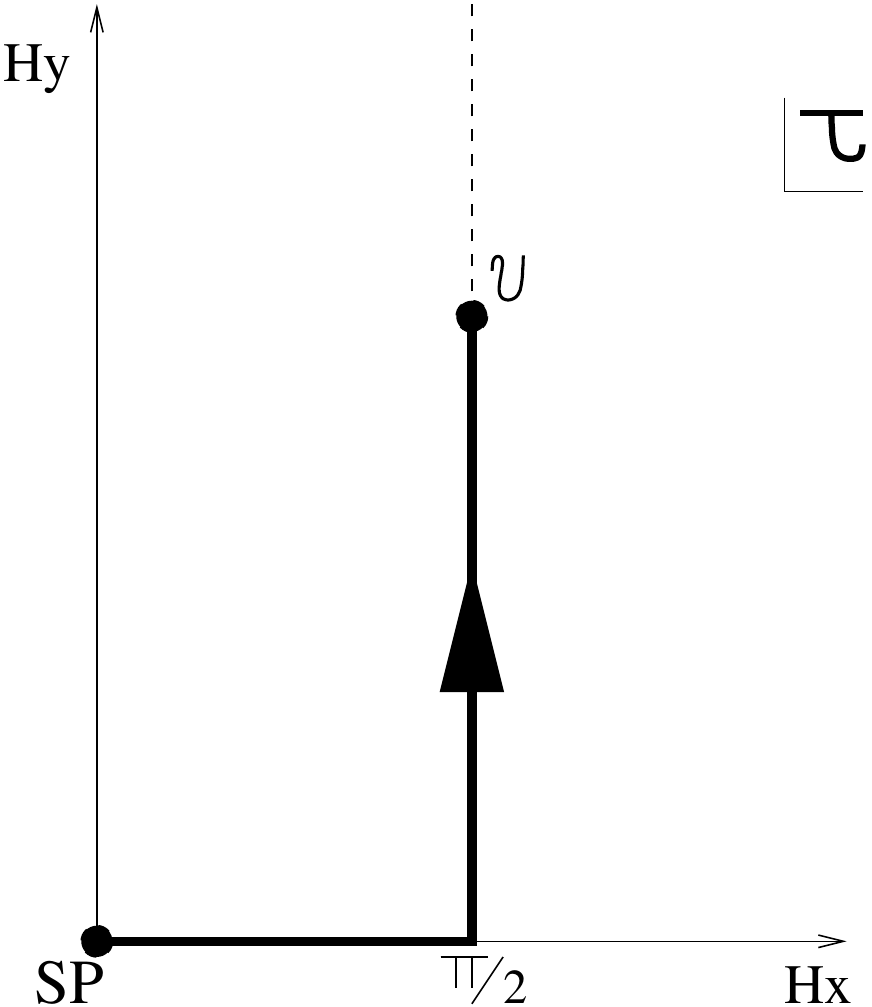}\hfill 
\includegraphics[width=3.0in]{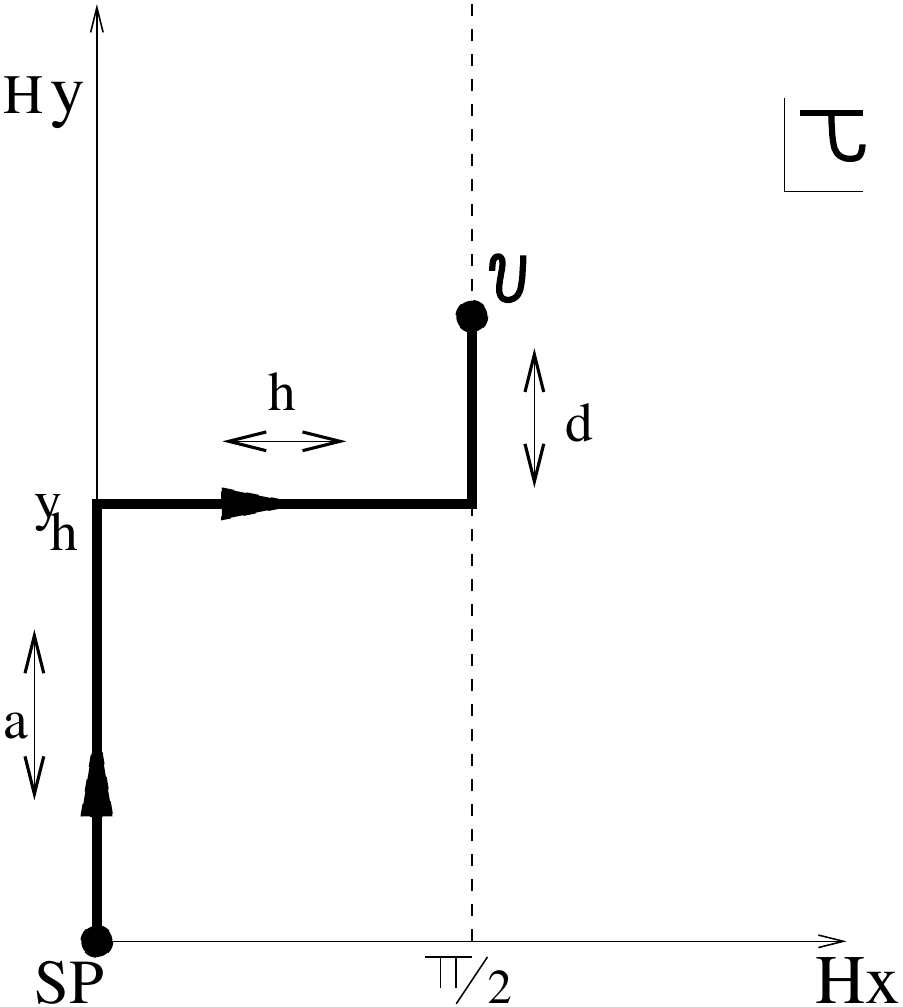} 
\caption{{\it Left panel}: The contour $C_D$ in the complex $\tau$-plane. The horizontal part is the geometry of half a Euclidean three-sphere. The vertical part is Lorentzian deSitter space (cf Fig \ref{iconic}). \\  {\it Right panel:} The contour $C_A$. The part (a) along the imaginary axis $x=0$ is AdS. The part (d) along the $x=\pi/2H$ line is Lorentzian deSitter space. The part (h) is a complex geometry that transitions between them. }
\label{contours}
 \end{figure*}

By choosing a particular contour connecting $\tau=0$ to $\tau=\upsilon = \pi/2H + iy_{\upsilon}$ we obtain a concrete representation of the geometry of the deSitter saddle point. The contour $C_D$ in Figure \ref{contours} gives its familiar representation. Along the part of $C_D$ from $x=0$ to $\pi/2H$ the geometry is the real Euclidean geometry of half a three-sphere. Along the part of $C_D$ from $(\pi/2H, 0)$ to $(\pi/2H, y_{\upsilon})$ the geometry is half of Lorentzian deSitter space,
\be
ds^2 = -dy^2 + \frac{1}{H^2}\cosh^2(Hy) d\Omega_3^2 .
\label{LordS}
\ee
The geometry along $C_D$  is often pictured by the iconic image in Figure \ref{iconic} (\lt).  

\subsection{An AdS Representation of a dS Saddle Point}

\begin{figure}[t]
\includegraphics[width=3.1in]{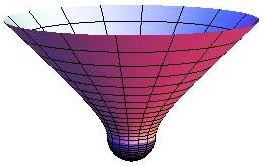}\hfill
\includegraphics[width=3.1in]{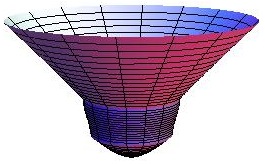}
\caption{Two embedding diagrams showing the different geometries representing the same deSitter saddle point. The generally complex geometries are embedded using the real metrics defined by the modulus of the scale factor $|a|$. The two figures  show the representation of a two dimensional slice of the same saddle point corresponding to an equator of the three-sphere in terms of the contours $C_D$  and $C_A$ in Figure \ref{contours}. The slices along the vertical parts of these contours are embedded in a flat Lorentz signatured  three-dimensional space. They are shaded in red. The slices along the horizontal parts of the contours are embedded in a Euclidean three dimensional flat space and shaded blue. The \lt figure is the NBWF deSitter saddle point as half a Euclidean three sphere joined to half a Lorentzian deSitter space across an equator. The next figure corresponds to the contour $C_A$ and consists of Euclidean AdS space joined (moving upwards) to the geometry of the horizontal branch, and then to deSitter space. Both representations give the same action and, in the more general case discussed below, the same predictions for the ensemble of classical histories.}
\label{iconic}
 \end{figure}
The contour $C_D$ is not the only useful representation of the dS saddle point. Consider the contour $C_A$ shown in the right panel of Figure \ref{contours}. This has the same endpoint $\upsilon$, the same action, and makes the same prediction for Lorentzian deSitter space as $C_D$. But the geometry is different.  The contour can be divided into a part (a) vertically along the imaginary $\tau$-axis to an intermediate point $\upsilon_a=iy_h$, a part (h) along the horizontal branch connecting $\upsilon_a$ to $\upsilon_b=\pi/2H+iy_h$, and finally a part (d) vertically along that line to the endpoint at $\upsilon$.

The geometry along part (a) is especially interesting. Evaluating the scale factor along $\tau=iy$ we get from \eqref{eucmetric} the line element
\be
ds^2 = -dy^2 - \frac{1}{H^2}\sinh^2(Hy) d\Omega_3^2 .
\label{EucAdS}
\ee
This is a negative signature representation of the geometry of Euclidean anti-deSitter space
with a cosmological constant $\tilde \Lambda = -\Lambda=-3H^2$. The geometry along the horizontal branch (h) is complex.  Finally along the $x=\pi/2H$ axis (d) it is real Lorentzian deSitter space as in \eqref{LordS}. 

{The fact that the dS saddle point has a representation in which the Euclidean regime is a negative signature AdS space with cosmological constant $-\Lambda$ follows immediately from the action \eqref{curvact}-\eqref{mattact}. Indeed with a negative signature convention, this is minus the action of Einstein gravity coupled to a negative cosmological constant $-\Lambda$ and a negative potential $-V$. At this point we remind the reader that we adopted a positive signature convention on the configuration space\footnote{This means in particular that the AdS saddle points with endpoint $\upsilon$ on the imaginary axis, where the boundary three-metrics are $h_{ij}=-(1/H^2)\sinh^2(Hy_h) \gamma_{ij}$, do not contribute to the NBWF as defined in \eqref{nbwf}.} $(h,\chi)$ of the NBWF. However this final boundary condition does not determine whether the interior  metric of the complex saddle points is positive or negative. Indeed a notion of signature is meaningless for complex  metrics.
By going along the imaginary axis with the contour $C_A$ the representation of the saddle point given here exhibits a negative signature Euclidean interior which, therefore, obeys the Euclidean AdS equations of motion following from \eqref{curvact}-\eqref{mattact}.

The Euclidean AdS regime of the saddle point is joined smoothly to a complex transition region. This represents the transition from a Euclidean to a Lorentzian geometry which in this representation involves a change in the signature of the boundary three-geometry rather than the $g_{\tau \tau}$ component of the metric. The saddle point geometry along $C_D$  is pictured in Figure \ref{iconic} (\rt).} The actions from the three parts of $C_A$  can be evaluated from the integral \eqref{eucact_tau}. The three contributions are:
\begin{subequations}
\label{actA}
\begin{align}
I_a(\upsilon_a)& = -\frac{\pi}{2 H^2}[1-\cosh^3(Hy_h)] ,
\label{Ia} \\
I_h(\upsilon_a,\upsilon_b) &= -\frac{\pi}{2H^2}[\cosh^3(Hy_h)-i\sinh^3(Hy_h)],
\label{Ih} \\
I_d(\upsilon_b,\upsilon) &= -\frac{\pi}{2H^2}[i\sinh^3(Hy_h)-i\sinh^3(Hy_{\upsilon})].
\label{Id}
\end{align}
\end{subequations} 
Evidently 
\be
I(\upsilon)=I_a(\upsilon_a) + I_h(\upsilon_a,\upsilon_b) + I_d(\upsilon_b,\upsilon) \ .
\label{actpts}
\ee
{As expected from the discussion above the contribution $I_a$ equals minus the usual action for Euclidean AdS bounded by a three-sphere of radius $y_h$ or, equivalently, a scale factor $(1/H)\sinh(Hy_h)$. }

\subsection{Regulation by Classicality}

We have seen that the real part of the action $I_R(y)$ remains constant as $y$ is increased along the $x=\pi/2H$ line [cf \eqref{dS}]. In general classical evolution requires that $I_R$ be constant along a Lorentzian history that is an integral curve of $S$.  That is necessary so that  $\exp(-2I_R)/\hbar$ can be the leading order probability of the history. In the present case, the Lorentzian history is at $x=\pi/2H$ so that $I_R(y)$ is constant [cf. \eqref{dS}]. 

By contrast, the contribution $I_a$ from along the vertical AdS part of the contour $C_A$ in \eqref{Ia} diverges with increasing $y_h$. That divergence however is cancelled in the action $I$ by the contribution $I_h$ from the horizontal part of the contour. In effect, classicality has regulated the action $I_a$. 

{\cf The divergence of the gravitational action in (asymptotically) Euclidean AdS spaces has been much discussed in the context of the AdS/CFT correspondence. There one considers AdS saddle points with endpoints on the imaginary axis in Fig \ref{contours} and one adds by hand a finite number of counterterms to the action in order to render it finite as the boundary is moved off to infinity \cite{Skenderis98,Skenderis02}. These counterterms can be expressed solely in terms of the boundary geometry $h_{ij}$. 
In four dimensions there are two gravitational counterterms, given by (with signs appropriate for the negative signature in \eqref{EucAdS})}
\begin{subequations}
\label{skcntr}
\begin{align}
I_1[h] &\equiv \frac{H}{4\pi}\int  d^3x \sqrt{-h}
 = \frac{\pi}{2H^2} \sinh^3(Hy_h) \ ,  \label{I1}   \\
I_2 [h] &\equiv \frac{1}{16\pi H} \int d^3x \sqrt{-h} \ {{^3}R(-h)} 
=\frac{3\pi}{4H^2} \sinh{Hy_h} . 
\label{I2}
\end{align}
\end{subequations}
where the last term in each equation is the counterterm evaluated in the three-metric of \eqref{EucAdS} at the value $\upsilon_a$. 
In terms of these counterterms, the contribution $I_a$ from along the vertical part of the contour $C_A$ is given by
\be
\label{regla}
I_a(\upsilon_a)  = -I^{\rm reg}_{AdS} + I_1(\upsilon_a) + I_2(\upsilon_a) +{\cal O}(e^{-Hy_h}) 
\ee
where $-I^{\rm reg}_{AdS}$ is the limiting value of $I_a - I_1 -I_2$ for $y_h \rightarrow \infty$. For the dS saddle point we consider here $I^{\rm reg}_{AdS} = \pi/2H^2$, which is the regularized action of Euclidean AdS space bounded by a scale factor $(1/H)\sinh{Hy_h}$. 

As anticipated, $I_a(y_h)$ exhibits the usual volume divergences of the Euclidean AdS action. However the real part of the horizontal contribution $I_h(y_h)$ is given by
\be
\Re [I_h(\upsilon_a,\upsilon_b)]  =- I_1(\upsilon_a) - I_2(\upsilon_a) +{\cal O}(e^{-Hy_h}) 
\label{regIh}
\ee
which supplies precisely the counterterms needed to regulate the volume divergences of $I_a$. Furthermore it does not contribute to the finite real part  $I_R(y_{\upsilon})$ of the saddle point action. Indeed, from \eqref{actpts}, \eqref{regla}, and \eqref{Id} we have, up to terms of order $\sim e^{-Hy_h}$,
\be
\Re [I(\upsilon)] \equiv I_R(\upsilon) = -I_{AdS}^{\rm reg}(\upsilon_a) \ .
\label{aprobs}
\ee
Thus we see that the probability of the Lorentzian deSitter history is given by the regulated action of the Euclidean AdS regime of the corresponding saddle point.

In the context of the AdS/CFT correspondence it has been argued that the counterterms correspond to expected renormalizations in the dual field theories (see e.g. \cite{Witten98}). However from a gravitational perspective their origin has remained somewhat obscure. In our framework the counterterms {are not added by hand. Instead they have a physical interpretation and arise automatically as part of the saddle point action when the latter} corresponds to a Lorentzian history. 

The classicality condition also implies that the wave function has a large phase factor $\exp(iS)$. This can be written in terms of the counterterms \eqref{skcntr} evaluated in the three-metric of \eqref{LordS} at the endpoint $\upsilon$. Hence one has,
\be
I(\upsilon) = -I^{\rm reg}_{AdS} + iS_{ct}(\upsilon)+{\cal O}(e^{-Hy_h}) 
\ee
where $iS_{ct} \equiv I_1(\upsilon)+I_2(\upsilon)$. We note that $S_{ct}$ is real, since $I_1(\upsilon_b)+I_2(\upsilon_b)=i(I_1(\upsilon_a)+I_2(\upsilon_a))$.

\section{Ensemble of Homogeneous Saddle Points}
\label{nonlinear}

In the previous section we obtained the following three results in the case of pure gravity and no matter for the dS saddle point represented by the contour $C_A$ in Figure \ref{contours}:  (1) The geometry along the vertical part of the contour (a) is asymptotically AdS for large radius $y$. (2) The real part of the action from the horizontal (h) part of the contour regulates the divergences of the AdS action. The imaginary part supplies the complex phase necessary for classicality. (3) The finite, non-divergent part of the action along the vertical part of the contour (a) supplies the probability for the Lorentzian deSitter history. 

When matter represented by a single scalar field is included the NBWF predicts a one-parameter family of saddle points that correspond to homogeneous isotropic classical histories \cite{HHH08}. In this section we show that these saddle points also admit a representation with a Euclidean AdS regime for which the same three properties hold. The analysis is qualitatively the same as in the previous section. The only difference is that we no longer have analytic solutions for the saddle point metric in the complex $\tau$-plane. However the asymptotic expansions for the solutions to the differential equations \eqref{euceqns} determining the saddle point suffice to obtain these results.

\subsection{Asymptotic expansions}

We will be interested in the behavior of the saddle point geometries for large $y=\Im(\tau)$, which is large radius in the Euclidean AdS regime, and where the $y_h$ defining the horizontal part of $C_A$ is appropriately located\footnote{{\qf For certain applications to cosmology it may be convenient to locate the horizontal part of the contour at reheating, and to evolve the universe classically from there onwards.}}. To that end we set $H=1$ just for this section, and introduce 
\be
u\equiv e^{i\tau} = e^{-y+ix} . 
\label{defu}
\ee
In this variable the saddle point equations \eqref{euceqns} become
\begin{subequations}
\label{eqnsu}
\be
-(ua')^2 -1 + a^2 + a^2 [(u\phi')^2 + 2V(\phi)]=0 \ ,
\label{eqnsua}
\ee
\be
u(u\phi')' +3 \frac{ua'}{a}(u\phi') + \frac{dV}{d\phi} =0 \ .
\label{eqnsuphi}
\ee
\end{subequations}
Here and elsewhere in this section $f'\equiv df/du$.  The key point is that all the coefficients in these equations are analytic in $u$. Therefore we can expect to find asymptotic expansions in powers of $u$ for the small values of $u$ when  $y$ becomes large. We verify this by explicit construction. In making these expansions we will assume that for small $\phi$, $V(\phi) =(1/2) m^2\phi^2 +{\cal O}(\phi^4)$, thus defining $m$. At large scale factor the field will have rolled down the hill and only this behavior of the potential near the minimum will be relevant for the leading terms in the asymptotic behavior. 
To keep the discussion manageable we will restrict attention to masses within the range $2<m^2<9/4$. The analysis is similar for smaller (positive) masses $m^2$ (and possibly fundamentally different for masses outside this range).

The general form of the asymptotic expansions has been worked out in \cite{Skenderis02} for Euclidean AdS spaces. Here we provide a  complex generalization of this. At large $y$ (small $u$) we expect the scale factor to be large and the field small. The leading behavior of the scale factor will be determined by the cosmological constant. From \eqref{eqnsua} we find 
\be
a(u)=\frac{c}{u}\left(1+\frac{u^2}{4c^2}\right)+ \cdots
\label{a_leading}
\ee
for some constant $c$. The constant $c$ can be adjusted to any value by translating $\tau$ by an appropriate amount [cf \eqref{defu}], that is, by changing the location of the SP in the complex $\tau$-plane. In the previous section we assumed that the SP was located at $\tau=0$ with the result that $c=i/2$ [cf. \eqref{defa}]. However, we have no access to regions near the SP in this asymptotic analysis. We therefore leave the constant $c$ undetermined. It has trivial physical content. 

Assuming that \eqref{a_leading} gives the leading asymptotic behavior for $a(u)$, we can calculate the form of the solution for $\phi(u)$ from \eqref{eqnsuphi}. The result is 
\be
\phi(u) = u^{\lm}(\alpha +\alpha_1 u + \cdots)  +  u^{\lp}(\beta +\beta_1 u +\cdots) 
\label{phia}
\ee
where
\be
\lp\equiv \frac{3}{2}[1+q(m)],  \quad \lm\equiv \frac{3}{2}[1-q(m)]
\label{ab}
\ee
with 
\be
q(m) \equiv \sqrt{1-(2m/3)^2}.
\label{defq}
\ee
Substituting the expansion  \eqref{phia} into \eqref{eqnsua} confirms the consistency of the ansatz \eqref{a_leading}--\eqref{phia} and determines the form of the next few terms in the asymptotic expansion of $a(u)$. We find
\be
a(u) =  \frac{c}{u}\left(1+\frac{u^2}{4c^2} +c_{-} u^{2\lambda_{-}} + c_3 u^3+ \cdots\right)
\label{a_series}
\ee
where $c_{-}$ and $c_3$ are further constants. 

Determination of all the coefficients in the expansions \eqref{phia} and \eqref{a_series} would require integrating the equations \eqref{euceqns} from the SP. The results would depend on the detailed physics of the matter (the shape of the potential) and on the NBWF conditions of regularity at the SP. Such integrations were carried out numerically in \cite{HHH08} for a quadratic potential. 

However, it follows from the equations \eqref{eqnsu} that the next coefficient $c_{-}$ in \eqref{a_series}  is fully determined in terms of the leading behavior. In particular we find
\be
c_{-} = -(3/4) \alpha^2
\ee
{\cf This is a generalization to complex geometries of a well-known result in asymptotically AdS spaces that the asymptotic solutions are locally determined from the asymptotic equations in terms of the `boundary values' $c^2\gamma_{ij}$ and $\alpha$, up to the $u^3$ term in \eqref{a_series} and to order $u^{\lambda_{+}^{\ }}$ in \eqref{phia}. 

Even though the next coefficients are not completely determined by the asymptotic equations, \eqref{eqnsu} still determines relations between them. In particular we find that for \eqref{phia} and \eqref{a_series} to satisfy \eqref{eqnsu} up to order $u^3$ the following relation must  hold: 
\be \label{coeff}
3c_3 +2m^2 \alpha \beta =0
\ee

\subsection{Asymptotic dS Saddle Points} 
\label{asympdS}

The asymptotic expansions derived in the previous subsection provide analytic information about the asymptotic form of the saddle points representing classical Lorentzian histories and about their probabilities.  We recall that Lorentzian histories correspond to curves in the complex $\tau$-plane along which both the scale factor and field are real and along which the classicality condition is satisfied. This means that  the real part of the action $I_R(\upsilon)$ varies slowly compared with the imaginary part $-S$ along the curve.

We already know what these curves are because we computed them numerically in \cite{HHH08}.  They are curves that are asymptotic to certain values  $x_r$ of constant $\Re(\tau)$. We found these curves by starting at the SP with a complex value of the scalar field $\phi(0)$. By tuning the phase of $\phi(0)$ together with the value of $x_r$ it was possible to find asymptotically vertical curves $x=x_r$  along which $a$ and $\phi$ were both real and the classicality condition satisfied. There was one such curve,  defining one classical Lorentzian history, for each value within a range of values of $\p0\equiv|\phi(0)|$. A one parameter family of homogeneous and isotropic, asymptotically deSitter, classical Lorentzian histories was thus predicted by the NBWF.

The asymptotic expansions allow us to identify the same curves partially analytically at large scale factor. 
To leading order in $u$ we have from \eqref{a_leading} and \eqref{phia}
\be
a(u) = \frac{c}{u}= |c|e^{i\theta_c} e^{-ix}e^y, \qquad \phi(u) = \alpha u^{\lambda_{-}^{\ }} = |\alpha|e^{i\theta_{\alpha}}e^{i\lambda_{-}^{\ }x}e^{-\lambda_{-}^{\ }y}
\label{lead_phase}
\ee
where $c$ and $\alpha$ are complex constants that are not determined by the asymptotic equations.
We are interested in solutions for which $a$ and $\phi$ are both real in the large $y$ limit along a constant value $x_r$. Hence for given $x_r$  the phases $\theta_c$ and $\theta_{\alpha}$ of such a solution are tuned so that
\be
\label{phases}
\theta_c=x_r, \qquad \theta_{\alpha}=-\lambda_{-}^{\ }x_r .
\ee
{\cf We emphasize it does not follow from the asymptotic analysis} that such tuning is possible with regularity conditions at the origin. However the numerical calculation\footnote{We cannot determine the values of $x_r, \theta_c,\theta_{\alpha}$ from the asymptotic analysis alone, but there is more information about them from the numerical calculations that we can briefly describe using the notation of \cite{HHH08}. In particular we can check whether the relations \eqref{phases} are satisfied for the values of $\mu$ consistent with the real $q$ assumed.  For small $\phi_0$ we must have $x_r \rightarrow \pi/2$ because this is what it is in the no matter case. That means that $\theta_{\alpha} \rightarrow - \lambda_{-}^{\ }\pi/2$ in this limit. For $\mu=3/4$ this means $\tan \theta_d \rightarrow -.32$, in agreement with the small $\phi_0$ limit of Fig 1(a) of \cite{HHH08}. 
Essentially the above relations predict  the curve $X(\phi_0)$ in Fig 1(b) of that paper once we know $\gamma (\phi_0)$.  Given that Fig1(b) shows that $x_r$ is significantly less than $\pi/2$ when $\phi_0$ is not small, this indicates  $x_a$  decreases for increasing $\phi_0$.
The consequence is that in the large $\phi_0$ limit in which $\theta_{\alpha} \rightarrow 0$ the Lorentzian slow roll vertical line approaches the imaginary axis and the AdS representation of the saddle point shifts to $x_a=-\pi/2$.}
 shows that it is for a range of $\p0$.

The asymptotic contribution to the saddle point action along the $x=x_r$ curves is given by the integral \eqref{eucact_alt} along the curve of constant $x=x_r$. It is immediate that there will be no contribution to the real part of the action $I_R$. The integrand is real,  but $d\tau=idy$. Thus
\be
\label{constIR}
I_R= \text{const}
\ee
when $x=x_r$ and large $y$.
The contribution from the asymptotic part of the contour is purely imaginary and $I_R$ remains constant thus satisfying the classicality condition. The curve $x=x_r$ is a Lorentzian history.

\subsection{AdS Domain Wall Representation of Asymptotic dS saddle points} 
\label{AdSDW}

The action of the asymptotic dS saddle points is given by the integral \eqref{eucact_alt} evaluated along a contour $C(0,\upsilon)$ connecting the SP to a point $\upsilon=x_r+iy_{\upsilon}$, with $y_{\upsilon}$ large. We now discuss the generalization of the contour $C_A$ employed in the AdS representation of the no-matter dS saddle point considered in the previous section.

\begin{figure}[t]
\includegraphics[height=4.in]{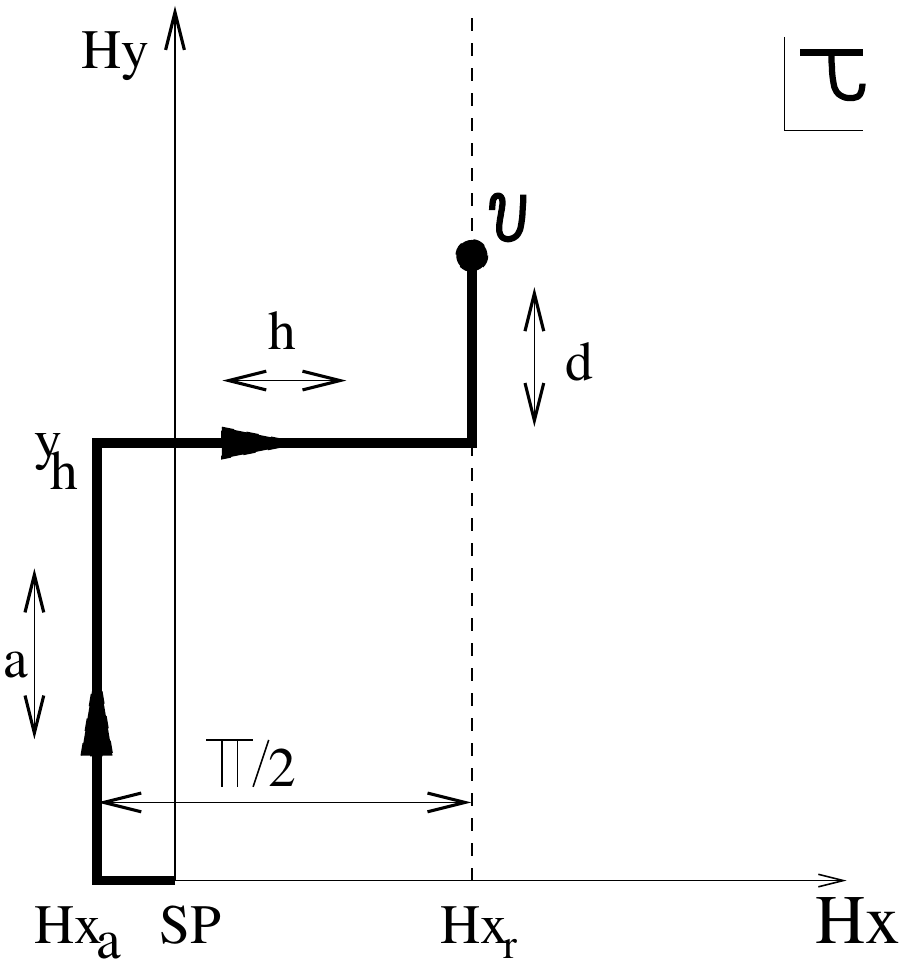}  
\caption{The contour $C_A$ when matter is included. The scale factor $a$ and field $\phi$ are real along an asymptotically vertical contour (d) but one which is displaced from $x=\pi/2H$ to $x=x_r$.  The vertical part (a) of the contour along which the saddle point is asymptotically AdS is shifted by an amount $\pi/2H$ in $x$ from that. The horizontal part of the contour (h) connects these two vertical parts --- relating AdS to dS.}  
\label{shifted}
 \end{figure}

Consider the vertical part at $x=x_a\equiv x_r - \pi/2$ of the contour shown in Figure \ref{shifted}. Eq.\eqref{defu} shows that the displacement from $x_r$ to $x_a$ replaces $u$ by $-iu$. The leading order behavior of $a(u)$ is replaced by $ia(u)$ [cf \eqref{lead_phase}]. Since $a$ was real along $x_r$ it will be imaginary along $x=x_a$.  The asymptotic form of the metric \eqref{eucmetric} along $x=x_a$ will then be 
\be
\label{adsmetric}
ds^2 = -dy^2 - (1/4)e^{2y} d\Omega_3^2 .
\ee
This is negative signature, real, Euclidean AdS. {\cf The asymptotic form of the scalar field along the $x=x_a$ curve is given by
\be
\phi(y) \approx |\alpha|e^{-i\lambda_{-}^{\ }\pi/2}e^{-\lambda_{-}^{\ }y}
\label{phase}
\ee
Hence the saddle point geometry along this part of the contour is that of an asymptotically AdS, spherically symmetric domain wall with a generally complex scalar field profile in the radial direction $y$. As before, since the domain wall has negative signature {\qf in our conventions} it is clear from the action \eqref{curvact}--\eqref{mattact} that it is a solution of Einstein gravity coupled to a negative cosmological constant $-\Lambda$ and a negative potential $-V$. The asymptotic phase of the scalar along the vertical part of the contour is locally (asymptotically) determined by the condition that it is asymptotically real along the $x=x_r$ curve. This means the phase factor in \eqref{phase} is universal, in the sense that it is independent of the dynamics and the regularity conditions in the interior. 

{\zf The generalization of the contour $C_A$ to include matter is therefore the one illustrated in Figure \ref{shifted}. There is a part (a) up the curve $x=x_a\equiv x_r-\pi/2$ to $\upsilon_a=x_a+iy_h$. There is a horizontal part (h) connecting this to $\upsilon_b = x_r+iy_h$ and then there is the part (d) up the $x=x_r$ axis to the endpoint $\upsilon$. (We continue to call this shifted contour $C_A$.) }

{\zf Of the results we obtained in the no-matter case mentioned at the start of this section, two generalizations to include scalar matter are immediate. (1) The vertical part of the contour (a) is along a curve where the geometry is asymptotically AdS. That was the content of \eqref{adsmetric}. The contribution to the action from this part of the contour is equal to minus the Euclidean AdS action of the domain wall solution and therefore exhibits the usual volume divergences of the AdS action. (2) The contribution to the saddle point action from the horizontal part (h) regulates the divergences from (a). This follows immediately from the fact that the real part of the action along (d) is constant [cf \eqref{constIR}]. The contribution from (h) therefore must cancel the divergences in (a).}

{\zf There remains the relation between the finite `regulated' action $I^{reg}_a= -I^{reg}_{DW}$ on (a), where  $I^{reg}_{DW}$ is the regulated Euclidean AdS action of the domain walls, and the saddle point action $I$ at $\upsilon$ on (d). This connection is supplied by the action integral \eqref{eucact_alt} along (h).  In the next section we show that for general, asymptotically inhomogeneous saddle points the horizontal branch (h) does not contribute to the finite part of the action at $\upsilon$.}

\section{General Saddle Points}
\label{more_general}
The results of the previous sections are not limited to homogeneous and isotropic models. They can be extended to more general saddle points without these symmetries. These are described by complex metrics
\begin{equation}
ds^2=N^2(\lambda) d\lambda^2 +g_{ij}(\lambda,\vx) dx^i dx^j
\label{eucmetric_hij}
\end{equation} 
where we use $\vx$ to indicate dependence on the three coordinates $(x^1,x^2,x^3)$ locating points on the compact spatial manifold. The complex variables $\tau$ and $u$ can be introduced as in \eqref{deftau} and \eqref{defu} respectively. A saddle point is a complex solution of the Einstein equations that is regular at the SP located by convention at $\tau=0$. Appendix \ref{hijstuff} gives a summary of the action and extremum equations in terms of the variables in \eqref{eucmetric_hij}. 

The asymptotic form of the general solutions of the Einstein equations for large $y$ (small $u$) has been worked out in detail (see e.g. \cite{Skenderis02}). Using these results,  we write for the expansion of the metric
\begin{subequations}
\label{expansions}
\be
\label{hijexpn}
g_{ij}(u,\vx)=\frac{c^2}{u^2}[\th_{ij}(\vx) +\th_{ij}^{(2)}(\vx) u^2 + \th_{ij}^{(-)}(\vx)u^{\lambda_{-}} +\th_{ij}^{(3)}(\vx)u^3 +\cdots] . 
\end{equation}
where $\th_{ij}(\vx)$ is {\zf real} and normalized to have unit volume thus determining the constant $c$. 
Then for the field
\be
\label{phia_hij}
\phi(u,\vx) = u^{\lm}(\alpha(\vx) +\alpha_1(\vx) u + \cdots)  +  u^{\lp}(\beta(\vx) +\beta_1(\vx) u +\cdots) . 
\ee
\end{subequations}
{\uf As in the homogeneous case, the asymptotic solutions are locally determined from the asymptotic equations in terms of the `boundary values' $c^2 \th_{ij}$ and $\alpha$, up to the $u^3$ term in \eqref{hijexpn} and to order $u^{\lambda_{+}^{\ }}$ in \eqref{phia_hij}.  Beyond this order the interior dynamics and the boundary condition of regularity on $M$ become important.}
 
We assume that along the contour $C_A$ in Fig \ref{shifted} the phases at the origin can be tuned so that $g_{ij}$ and $\phi$ are real along the vertical\footnote{It follows directly from the expansions \eqref{hijexpn} and \eqref{phia_hij} that if a given saddle point corresponds to a Lorentzian history the latter must lie on an asymptotically vertical curve in the complex $\tau$-plane.} part (d) for small $u$. {\zf Otherwise the saddle point does not correspond to a Lorentzian history and is suppressed in the path integral.} Since the expansions are analytic functions of $u$ that means that there is a parallel contour (a) at $x_a=x_r-\pi/2H$ along which the metric $g_{ij}$ is also real but with the opposite signature. Thus we recover more generally the same story as in the homogeneous and isotropic example. 

It remains to establish the connection between the contribution to the finite part of the action from the vertical part of the contour $C_A$ and the saddle point action at the endpoint $\upsilon$. To this end we calculate the action integral along the horizontal branch (h) of the contour  connecting (a) to (d) order by order in $u$. Using the expansions \eqref{hijexpn} and \eqref{phia_hij} we find
\be
\label{horizhij}
I_h(\upsilon_a,\upsilon_b)=\frac{1}{8\pi}\int_{x_a}^{x_r}dx\int d^3x \ g^{\frac{1}{2}} \left[6H^2 -{^3}R + 2V(\phi) +6({\vec\nabla}\phi)^2 \right]
\ee
where ${^3}R$ is the scalar three curvature of $g_{ij}$. 
In Appendix \ref{hijstuff} we show that this does not contribute to the finite part of the action in the large $y_h$ limit as a consequence of the asymptotic Einstein equations. This means $I_h$ only regulates the divergences of the action from (a) and supplies phase\footnote{It does not supply all of the phase as it did in the no-matter example because the field is generally complex along (a) which renders $I^{reg}_a$ complex.} necessary for classicality on (d). In particular we obtain
\be
\label{horizhij2}
I_h(\upsilon_a,\upsilon_b) = (i-1)(I_1 +I_2+I_3)(\upsilon_a)+{\cal O}(e^{-y_h})
\ee
where $I_1$ and $I_2$ are the familiar (real) gravitational counterterms and $I_3  \sim \int \sqrt{h}\phi^2$ is an additional (complex) scalar field counterterm \cite{Skenderis02} which cancels the $e^{3q}$ divergence arising from the slow fall-off of $\phi$ for large $y_h$. In the range of values of $q$ we consider here the scalar gradient term in \eqref{horizhij} decays at large scale factor.  {\clr Hence for sufficiently large $y_h$ the combination 
\be
I_a(\upsilon_a) -(I_1+I_2+I_3)(\upsilon_a) 
\label{reg}
\end{equation}
approaches a finite asymptotic limit which is the so-called regulated Euclidean AdS action of the domain wall, $-I^{\rm reg}_{DW}$. Finally, the contribution from the (d) part of the contour is purely imaginary and given by
\be
\label{horizhij2}
I_d(\upsilon,\upsilon_b) = i(I_1 +I_2+I_3)(\upsilon)- i(I_1 +I_2+I_3)(\upsilon_b)+{\cal O}(e^{-y_h})
\ee

Hence the sum of the actions from the three parts of the contour $C_A$ express the action of a general, inhomogeneous saddle point of the NBWF at $\upsilon=x_r +iy_\upsilon$ for sufficiently large $y_h$ and $y_\upsilon$ in terms of the regulated domain wall action and a sum of purely imaginary surface terms, 
\be
I [\upsilon,\th_{ij}(\vx),\chi(\vx)] = -I^{\rm reg}_{DW} [\th_{ij}(\vx),\alpha(\vx)] + iS_{ct}[\upsilon, \th_{ij}(\vx),\alpha(\vx)] +  {\cal O}(e^{-y_\upsilon})
\label{scnbwf}
\ee
where $iS_{ct} \equiv (I_1+I_2+I_3)(\upsilon)$. 

In these equations $\alpha$ is locally (asymptotically) determined by the argument of the wave function as described in Section \ref{nonlinear}. Thus we find that in the limit of large scale factor} the probabilities for general perturbed, Lorentzian, asymptotically deSitter histories with scalar matter can be found from the Euclidean AdS domain wall regime of the saddle point. 

{\qf Eq. \eqref{scnbwf} relies on the validity of the asymptotic expansions \eqref{hijexpn} and \eqref{phia_hij}. It is therefore implicitly assumed in \eqref{scnbwf} that for a given set of boundary values $(\th_{ij}(\vx),\alpha(\vx))$, the `scale factor' $\sim 1/u$ all along the horizontal branch of the contour is sufficiently large for the expansions to hold. However if one considers the wave function on a surface $\Sigma$ at a given, finite value of $b$ then there always exist configurations $(\th_{ij},\chi)$ for which the expansions \eqref{hijexpn}--\eqref{phia_hij} don't hold near $\Sigma$. The saddle point action of such configurations is not given by the regularized asymptotic domain wall action $I^{reg}_{DW}$. In fact, the condition that the asymptotic expansions hold across the horizontal branch is closely related to the requirement discussed above that the configuration behaves classically near $\Sigma$.

Classicality requires a certain coarse-graining over the fine details of the geometry and matter field configuration on $\Sigma$. This introduces a length scale $\zeta$.} In general the saddle point action of a coarse-grained configuration will depend on $\zeta$. Hence at a given, finite scale factor $b$ we have, {\clr from \eqref{scnbwf}},
\be
I [b,\th_{ij}(\vx),\chi(\vx)] = -I^{\rm reg}_{DW} [\th_{ij}(\vx),\alpha(\vx),\zeta(b)] + iS_{ct}[b,\th_{ij}(\vx),\alpha(\vx)] +{\cal O}(e^{-y_\upsilon})
\label{scnbwf2}
\ee
where the functions $\th_{ij}(\vx)$, $\chi(\vx)$ and $\alpha(\vx)$ should be understood as labels for coarse-grained configurations that satisfy the classicality conditions on $\Sigma$. {\qf This ensures the expansions \eqref{hijexpn}--\eqref{phia_hij} hold in the transition region from AdS to dS which in turn implies that the regularized AdS action of the coarse-grained configuration at $b$ is approximately given by its asymptotic value $I^{reg}_{DW}$ {\it given} the coarse-graining scale $\zeta$.}}

{\uf An illustrative case to which \eqref{scnbwf2} applies is the wave function of linear fluctuations\footnote{We refer the reader to \cite{Maldacena03,HHH10a,Harlow11} for a more detailed discussion of the wave function of linear fluctuations around empty de Sitter space and Euclidean AdS space in the context of dS/CFT.} around homogeneous isotropic inflationary backgrounds \cite{HHH10a}. Eq.\eqref{scnbwf2} implies that the Gaussian wave function of fluctuations in inflationary universes can be obtained from a Euclidean AdS calculation by evaluating the action of perturbations around AdS domain walls.} 
{\qf Now, it is well known that only perturbation modes with wavenumber $n \leq bH$ evolve classically near a spacelike surface with scale factor $b$. Perturbation modes with shorter wavelengths are therefore summed over in the probabilities for the ensemble of classical configurations on $\Sigma$. In the context of perturbation theory therefore the coarse-graining scale $\zeta \sim 1/bH$ that enters in \eqref{scnbwf2} is associated with the horizon size $H^{-1}$.}

{\qf Given that $H^{-1}$ remains approximately constant during inflation, it follows that $\zeta$ is inversely proportional to $b$. The classical ensemble of the NBWF at smaller values of the scale factor emerges thus as a coarse graining of the ensemble at larger values of $b$. In the limit $b \rightarrow \infty$ one obtains the {\vf maximally refined ensemble of histories consistent with classicality.} }

\section{Holographic no-boundary Measure}
\label{dualities}

The AdS domain wall representation of the saddle points {\uf provides} a natural connection between the no-boundary amplitude of coarse-grained classical configurations and the Euclidean AdS/CFT duality. {This leads to a dual formulation of the semiclassical NBWF in this regime and hence a realization of a dS/CFT duality that we now describe.}

\subsection{AdS/CFT}

The Euclidean AdS/CFT correspondence \cite{Maldacena98,Witten98,Gubser98} postulates {\uf a relation between} the semiclassical supergravity partition function $Z_{SG}[\xi]$ in asymptotically AdS spaces and the large $N$ limit of the partition function $Z_{CFT}[\xi]$ of a dual conformal field theory defined on the conformal boundary. The boundary conditions $\xi$ on the dynamical fields in the gravity theory enter as sources in the dual partition function. 

The AdS/CFT duality is a strong/weak coupling duality, in the sense that the relation between the parameters in both theories is such that when the low energy gravity approximation can be trusted the CFT is strongly coupled {\uf (and at large $N$)} and vice versa. 
The duality therefore provides a powerful alternative way to compute CFT correlation functions using AdS gravity. To do this, one differentiates $Z_{CFT} \approx \exp (-I^{reg}_{SG}/\hbar)$ with respect to $\xi$, where $I^{reg}_{SG}$ is the regularized Euclidean action of a solution of the classical supergravity equations with asymptotic boundary conditions $\xi$ and smooth `no-boundary' conditions in the interior. 

An interesting class of supergravity solutions for which the duality has been explored in great detail concerns the real {\af versions} of the Euclidean AdS domain walls discussed above. These are regular Euclidean solutions, involving only gravity and a scalar field with a negative potential $-V$, in which the scalar field has a nontrivial profile in the radial AdS direction\footnote{The Lorentzian continuation of domain walls of this kind describe collapsing cosmologies that produce a big crunch singularity in AdS \cite{SING}.}. Explicit examples were found e.g. in the consistent truncations of ${\cal N}=8$ gauge supergravity in four dimensions obtained in \cite{Duff99}, for which the dual theory is known \cite{Aharony08}. In these models, the scalar potential $-V$ has a negative maximum around which the scalars have mass $M^2\equiv -m^2 =-2H^2$. 

The metric and scalar field in general AdS domain wall solutions behave asymptotically as \eqref{hijexpn} and \eqref{phia_hij}, with $\lambda_{\pm}^{\ }$ given by \eqref{ab}. {\uf In the supergravity limit} the AdS/CFT dictionary then states that
\be
\exp (-I^{reg}_{DW}[\tilde h_{ij},\alpha]/\hbar) = Z_{QFT}[\tilde h_{ij},\alpha] = \langle \exp \int d^3x \sqrt{\tilde h} \alpha {\cal O} \rangle_{QFT}
\label{operator}
\ee
{\uf where the dual QFT lives on the conformal boundary of AdS represented here by the three-metric $\th_{ij}$. On the right hand side the brackets $\langle \cdots \rangle$ denote the functional integral average involving the boundary field theory action minimally coupled to the metric conformal structure represented by $\tilde h_{ij}$.} For radial domain walls this is the round three-sphere, but in general $\alpha, \beta$ and $\tilde h_{ij}$ are functions of the boundary coordinates $x$. 

{\uf The AdS/CFT duality \eqref{operator} relates the asymptotic AdS factor $\exp (-I^{reg}_{DW}/\hbar)$ to the partition function of a dual field theory. The duality on a surface at finite radius $a$ emerges as a particular coarse-graining of \eqref{operator} {\qf over short scale degrees of freedom} on both sides. On the gravity side the coarse-graining is closely connected to that needed to ensure the expansions \eqref{hijexpn}--\eqref{phia_hij} hold near the boundary. This introduces a scale $\zeta \sim 1/a$ as discussed in Section \ref{more_general}. The regularized AdS action $I^{reg}_{DW}[\tilde h_{ij},\alpha,\zeta]$ that enters in the duality at finite radius is the asymptotic limit of the action of the coarse-grained configurations on the boundary and generally depends on $\zeta$.

On the field theory side the coarse-graining scale $\zeta$ defines a UV cutoff $\epsilon \sim 1/a$ that specifies the range of high-energy modes that should be integrated out, yielding a new partition function $Z_{QFT}[\tilde h_{ij},\alpha,\epsilon]$. Thus the dependence of the bulk wave function for coarse-grained configurations on the radius of the boundary emerges from the energy scale at which one considers the dual field theory. However, the precise mapping between the radius $a$ of the boundary and the UV cutoff $\epsilon$ in the dual has yet to be identified (see e.g. \cite{Heemskerk10,Faulkner11} for recent work on this). }

It follows from \eqref{operator} that AdS/CFT relates the ensemble of domain wall solutions of the action of a given (super)gravity theory to an ensemble of dual QFTs that are {\af deformations} of a given CFT. The {\af additions to the field theory action are} relevant deformations of the form $\alpha {\cal O}$, where the dual operator ${\cal O}$ has dimension $\Delta=\lambda_{+}^{\ }$. The deformation leads to a vev for ${\cal O}$ that can be computed from the gravity theory and that is given by the coefficient $\beta$ in \eqref{phia_hij}. The precise relation between the strength of the deformation $\alpha$ and the vev $\beta$ depends on the details of the theory and on the regularity condition in the interior \cite{SING}.
For example for the consistent truncations of \cite{Duff99} mentioned earlier the dual operator has dimension $\Delta=2$, and radial domain walls are associated with relevant deformations of ABJM theory  \cite{Aharony08} defined on the unit three sphere. 

The connection exhibited in \eqref{operator} between bulk boundary conditions $\xi$ and deformations of the dual field theory is a general feature of AdS/CFT that applies to all bulk fields. {\uf In the application of the correspondence to cosmology that we are about to describe we will regard} the partition function $Z_{QFT}[\xi]$ of the boundary theory as a function of these boundary values $\xi=(\alpha,\tilde h,...)$. }

\subsection{dS/CFT}

\begin{figure}[t]
\includegraphics[height=3.in]{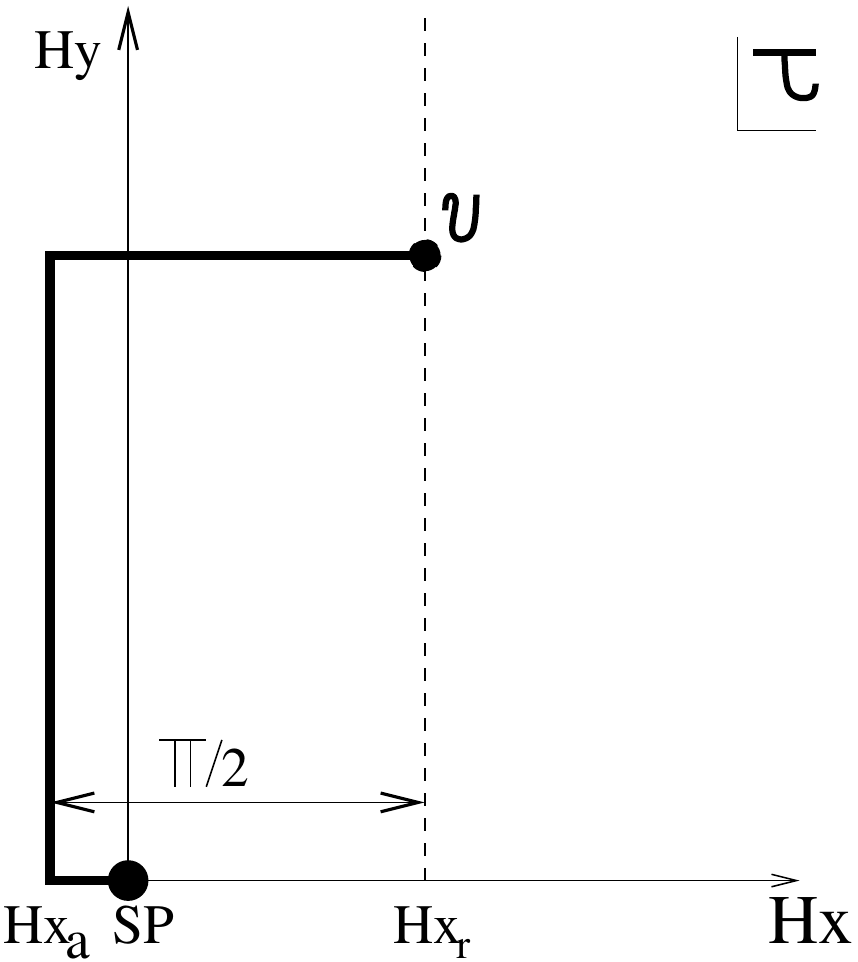}  
\caption{The saddle point representation that serves as a guide in the derivation of a dual formulation of the NBWF in terms of a field theory defined on the conformal boundary geometry at $\upsilon$.}  
\label{shifted2}
 \end{figure}

To derive a dual formulation of the semiclassical NBWF we consider a representation of the saddle points along contours of the kind shown in Fig \ref{shifted2}. We concentrate on models where $-V$ is a scalar potential of the form usually considered in AdS/CFT. That is, we consider potentials $-V$ for which the AdS theory is stable, so that the AdS/CFT dual is well-defined\footnote{This includes the requirement that the field satisfies the Breitenlohner-Freedman (BF) bound $m^2_{BF}$ on the scalar mass in AdS. Intriguingly we found in \cite{HHH08} that for scalars with masses $m^2>-m^2_{BF}$ the NBWF predicts no classical, homogeneous isotropic histories in a neighborhood around the empty deSitter history. Possibly related to this, it has also been suggested that scalar particles with masses in this range are unstable in deSitter space \cite{Bros06}.}. Whether the resulting dS/CFT duality can be generalized to models outside this class is an important open question with potentially important phenomenological implications which we do not address here.

We have shown (cf \eqref{scnbwf2}) that the leading order NBWF of coarse-grained classical configurations $(b,  \tilde h,\chi)$ is given by the product of an AdS factor $\exp(+I^{reg}_{DW})$, multiplied by a surface term that is the exponential of a phase factor $iS_{ct}$. The AdS factor governs the probabilities of different configurations on $\Sigma$ whereas the phase factor is important in predicting the classical evolution of the configuration. The action $I^{reg}_{DW}$ is the regularized action of coarse-grained Euclidean AdS domain walls with boundary conditions $(\tilde h,\alpha)$ where $\alpha$ is locally related to $\chi$. The coarse-graining ensures that $I^{reg}_{DW}$ reaches its asymptotic limit near $\Sigma$. Hence provided we consider a model with a known AdS/CFT dual we can apply the finite radius (or coarse-grained) version of the AdS/CFT relation \eqref{operator} in the (super)gravity limit to \eqref{scnbwf2}. This yields\footnote{This relation depends on the number of dimensions. See also \cite{Maldacena11}.} 
\be
\label{dscft}
\Psi [b,  \tilde h,\chi] = \frac{1}{Z_{QFT}[\tilde h, \alpha,\epsilon] }\exp(iS_{ct}[b,\chi, \tilde h]/\hbar)
\ee
where $\epsilon\sim 1/Hb$ is the UV cutoff in the dual field theory mentioned earlier. The sources $(\tilde h, \alpha)$ of $Z_{QFT}$ are associated with the asymptotic behavior along the AdS part of the contour in Fig \ref{shifted2} but they are locally related\footnote{That is, by the asymptotic equations only.} to the argument $(b,\tilde h,\chi)$ of the wave function at the dS endpoint $\upsilon$.

The dependence of the field theory partition function on the argument of the wave function gives a measure on different classical configurations on $\Sigma$. For sufficiently small values of the matter sources and sufficiently mild deformation of the round three sphere geometry one expects the integral defining the partition function to converge. Thus \eqref{dscft} gives a dual formulation in which the semiclassical no-boundary measure on classical configurations is given in terms of partition functions of field theories on the final boundary that are relevant deformations of the CFTs that occur in AdS/CFT. In models where the AdS/CFT dual is known explicitly this yields a concrete realization of a dS/CFT duality that in principle provides an alternative way to compute the cosmological probability measure in the no-boundary state\footnote{The complete dual field theory lives on the future conformal boundary $\tilde h$. Nevertheless it takes into account the initial state of the universe, because AdS/CFT implements a regularity condition in the interior of the saddle point. Indeed the Euclidean AdS/CFT correspondence is perfectly consistent with the notion of a unique no-boundary quantum state of the universe \cite{Horowitz04}.}. 

The evolution in time of the universe (as represented by changes in the scale factor) emerges as inverse RG flow in the dual theory as originally conjectured in \cite{Strominger01b} and more recently discussed in a different context in \cite{Bousso10,Vilenkin11}. In our setup this is because the scale factor $b$ on $\Sigma$ specifies the radius of the boundary of the AdS regime of the corresponding saddle point, which via AdS/CFT is related to a cutoff energy scale $\epsilon^{-1}$ in the dual field theory. Hence when one considers the NBWF on a surface $\Sigma$ at finite scale factor $b$, high energy modes in the dual field theory are coarse-grained over. We have argued this is just what one expects, both from the holographic principle in general and from the classicality conditions on $\Sigma$ in particular, which require a similar coarse-graining in the bulk. This can be made explicit in perturbation theory in inflationary cosmologies where classicality on $\Sigma$ requires a coarse-graining over subhorizon modes with wavenumber $n > \epsilon^{-1}$. When the scale factor increases, more modes contribute to the classical ensemble both in the bulk and in the boundary theory. In the asymptotic limit one obtains the {\vf maximally refined ensemble of histories consistent with classicality.

{\qf From a cosmological point of view it is natural to consider the amplitude of coarse-grained classical configurations. It is intriguing however that the coarse-graining associated with classicality in the bulk appears to be an intrinsic part of the AdS/CFT duality itself\footnote{It seems plausible this is also connected to the absence of precise local observables in quantum gravity.}. This suggests that an AdS/CFT dual description -- which captures well-defined asymptotic observables in quantum gravity -- provides a truly coarse-grained description of the bulk. It is plausible this is connected to the difficulties to decipher the physics of Hawking radiation from a dual perspective. Perhaps the details of this process are coarse-grained over in the dual description.}

It is natural to conjecture that the dS/CFT duality \eqref{dscft} extends beyond the leading order approximation.
This would place the no-boundary wave function on firm footing in string theory and in particular identify an asymptotic or coarse-grained regime in which it has a precise meaning. It would also open up the possibility to use the dual partition function at finite $N$ to compute the string and quantum corrections to the no-boundary measure. In fact, the higher spin realization of dS/CFT found recently \cite{Anninos11} provides some support for an exact duality of the form proposed in \eqref{dscft}. It would be interesting to explore this further e.g. by comparing the next order in $\hbar$ corrections\footnote{This is not an AdS/CFT calculation in the usual sense, since the AdS action enters with a plus sign in the gravitational path integral. This implies that fluctuations are damped which suggests the determinants arising from the next order corrections will be well-defined in the setup considered here.} on both sides of \eqref{dscft}.

\vskip .2in

\noindent{\bf Acknowledgments:} } {\af Many of the ideas in this work originated in discussions  with Stephen Hawking and we thank him for those.} We thank Dionysios Anninos, Frederik Denef, Gary Horowitz, Juan Maldacena, Paul McFadden, Joe Polchinski, Harvey Reall, Kostas Skenderis, David Tong, Antoine Van Proeyen and the participants of the Quantum gravity Institute at CERN for helpful conversations. We thank Chris Pope, Sheridan Lorenz and the Mitchell Institute for several meetings at Cooks Branch. The work of JH was supported in part by the US NSF grant PHY08-55415 and  by Joe Alibrandi. The work of TH was supported in part by the Agence Nationale de la Recherche (France) under grant ANR-09-BLAN-0157 and by the FWO-Vlaanderen under the Odysseus program.

\appendix

\section{The Action on the Horizontal Branch (h) of Contour $C_A$}
\label{hijstuff}
This appendix gives a few more details on the derivation of the form of the action $I_h(y_h)$  \eqref{horizhij}  on the horizontal branch ($h$) of the contour $C_A$. In particular we sketch a derivation of the key result that the finite contribution vanishes in an expansion of $I_h(y_h)$ for large $y_h$.  

We begin by writing the total action $I[g,\phi]\equiv I_C[g] +I_\phi[g,\phi]$ [cf \eqref{curvact},\eqref{mattact}] in standard 3+1 form for metrics of the form 
\be
ds^2=N^2(\lambda,\vx) d\lambda^2 +g_{ij}(\lambda,\vx) dx^i dx^j .
\label{eucmetric_hij-1}
\end{equation} 
We find
\be
I=\frac{1}{16\pi} \int d^4 x \left(\frac{1}{N}{\cal K} + N{\cal P}\right) .
\label{action}
\ee
Here, ${\cP N}$ is the `potential' part of the Lagrangian
\be
N{\cP}\equiv N g^{1/2}\left(6H^2-{^3}R +12 V +6(\vec\nabla\phi)^2\right)
\label{potential}
\ee
and $\cK/N$ is the `kinetic' part 
\be
\frac{1}{N} {\cal K} \equiv N g^{1/2}\left[K_{ij}K^{ij} - K^2 +6\left(\frac{\phi'}{N}\right)^2\right]
\label{kinetic}
\ee
with the extrinsic curvature defined by (no shift in \eqref{eucmetric_hij-1})
\be
K_{ij}\equiv \frac{1}{2N}\frac{\partial g_{ij}}{\partial\lambda} .
\label{kij_def}
\ee
Varying the action with respect fo $N(t,\vx)$ gives the (Hamiltonian) constraint equation
\be
\frac{1}{N}\cK = N\cP .
\label{constraint}
\ee
The value of the action on any solution of this constraint can be written
\be
I=\frac{1}{16\pi}\int d^4x \ 2N\cP = \frac{1}{16\pi}\int d^4x \ 2\cK/N .
\label{valact}
\ee
In these expressions $d^4x=d\lambda d^3x$. But by restricting to $N$'s that are a function of $\lambda$ only as assumed in \eqref{eucmetric_hij} we can rewrite our equations in terms of the complex parameter $\tau=x+iy$ introduced by \eqref{deftau}. Then on the horizontal branch ($h$) of $C_A$ we have\footnote{We apologize for having used `x' for so many different things.}
\be
I_h(y_h) = \frac{1}{8\pi} \int^{x_r}_{x_a} dx \int d^3x \cP =  \frac{1}{8\pi} \int^{x_r}_{x_a} dx \int d^3x \cK
\label{horizb}
\end{equation}
where now $K_{ij}\equiv (1/2)(\partial g_{ij}/\partial\tau)$.  The first of these expressions was quoted in \eqref{horizhij} because there is a more direct connection with the standard counter terms in \cite{Skenderis98}. But we could have used the $\cK$ form. 

The asymptotic form of $I_h(y_h)$ for large $y_h$ can be found by expanding $\cP$ and $\cK$ for large $y$ using the variable $u$ defined in \eqref{defu} and the expansions \eqref{hijexpn} and \eqref{phia_hij} derived in \cite{Skenderis98}. There are terms that diverge for small $u$. But for the finite part independent of u, we find 
\be
\left[\cP\right]_{\rm finite} = \frac{1}{16\pi} \sqrt{c^6\th}\left(3\th^{ij}\th^{(3)}_{ij} -12 m^2 \alpha(\vx)\beta(\vx)\right)=-[\cK]_{\rm finite}
\label{finiteP}
\ee
The constraint equation \eqref{constraint} can also be expanded in powers of $u$ with the result that 
\be
[\cP]_{\rm finite} = [\cK]_{\rm finite} . 
\label{finite_constr}
\ee
The implication of \eqref{finite_constr} and \eqref{finiteP} is that the finite part of the expansion of the action vanishes. 
\be
[\cP]_{\rm finite} = [\cK]_{\rm finite}=0
\label{finite_vanish}
\ee
It might seem that the vanishing of the finite part of the action on the horizontal contour is a consequence of the constraint equation \eqref{constraint} alone. But, in fact, the form of the expansions \eqref{hijexpn} and \eqref{phia_hij} rely on the other field equations as well, and the scalar field equation in particular.

\end{document}